\documentclass[a4paper]{jpconf}
\usepackage{graphicx}
\usepackage{amssymb,amsmath}
\usepackage{bm}

\begin{document}
\title{Rao-Blackwellised Interacting Markov Chain Monte Carlo for Electromagnetic Scattering Inversion}

\author{F. Giraud$^1$, P. Minvielle$^1$, 
M. Sancandi$^1$, P. Del Moral$^{2,}$}
\address{$^1$ CEA-CESTA, 33114 Le Barp, France}
\address{$^2$ INRIA Bordeaux Sud-Ouest, Domaine Universitaire, 351, cours de la Liberation, 33405 Talence Cedex, France}
\ead{francois.giraud@cea.fr}

\begin{abstract}
The following electromagnetism (EM) inverse problem is addressed. It consists in estimating local radioelectric properties of materials recovering an object from the global EM scattering measurement, at various incidences and wave frequencies. This large scale ill-posed inverse problem is explored by an intensive exploitation of an efficient 2D Maxwell solver, distributed on High Performance Computing (HPC) machines. Applied to a large training data set, a statistical analysis reduces the problem to a simpler probabilistic metamodel, on which Bayesian inference can be performed.  Considering the radioelectric properties  as a dynamic stochastic process, evolving in function of the frequency, it is shown how advanced Markov Chain Monte Carlo methods, called Sequential Monte Carlo (SMC) or  interacting particles, can provide estimations of the  EM properties of each material, and their associated uncertainties.
\end{abstract}

\section{Introduction}
\begin{figure}[ht!]
\centerline{\includegraphics[width=0.6\columnwidth]{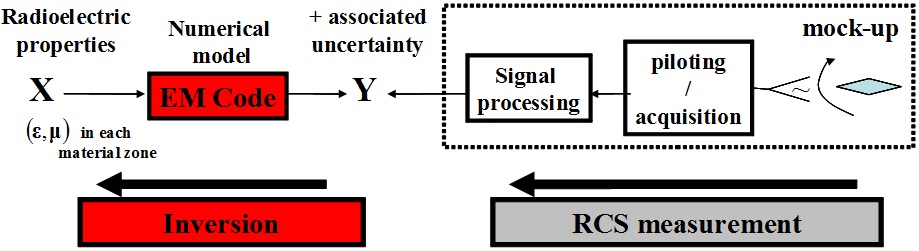}}
\caption{The EM inverse problem}
\label{PbInv}
\end{figure}

The inverse problem is described in figure \ref{PbInv}. The Radar Cross Section quantifies the scattering power of an object, at a given incidence and wave frequency. It is defined as the ratio between the radar transmitted power and the incident power density (in plane wave) \cite{Knott}. RCS measurement process is schematically presented on the right part of figure \ref{PbInv}. The object or mock-up is illuminated by a quasi-planar monochromatic wave, inside an anechoic chamber where interferences are limited.The acquisitions are realized at $K$ successive discrete frequencies (${f_1,f_2,\cdots,f_K}$), for different incidence angles (by piloting  motorized rotating support). From this raw data, a signal processing is performed, mainly consisting of calibration and filtering. At the end, the measurement provides an evaluation of the calibrated complex (amplitude and phase) scattering coefficient, for each frequency and  incidence.

A metallic axi-symmetric object is recovered with $N$ areas (see figure \ref{freq-fix}), each area corresponding to a material with its associated isotropic radioelectric properties, i.e. the complex parameters of permittivity  $\epsilon$ and permeability $\mu$. The EM  inverse problem  can be expressed as: is it possible to extract some local information on the material properties $(\epsilon_k,\mu_k)$ of each area from the global scattering measurement?

\section{The stochastic modeling}
At a given frequency $f_k$, the system state $\mathbf{X}_k$  can be defined by (omitting $f_k$ to lighten the notations):
$
\mathbf{X}_k=\left[  \bm{\underline{\epsilon}}^{\prime} \text{   } \bm{\underline{\epsilon}}^{\prime\prime} \text{   } \bm{\underline{\mu}}^{\prime} \text{   } \bm{\underline{\mu}}^{\prime\prime} \right]^T
$,
 with $\bm{\underline{\epsilon}}^{\prime}$, $\bm{\underline{\epsilon}}^{\prime\prime}$, $\bm{\underline{\mu}}^{\prime}$ and $\bm{\underline{\mu}}^{\prime\prime}$, respectively  the real and imaginary permittivity and permeability components of the $N$ areas. Notice that the state dimension can be high ($4\cdot N$ with $N\approx 100$). On the other hand, considering measurements at the given frequency $f_k$ with various incidence angles in both polarizations (TM and TE), the observation vector is made of the real ($\Re(\cdot)$) and imaginary  ($\Im(\cdot)$) parts of the complex scattering coefficients $\mathbf{c_{TM}}$ and $\mathbf{c_{TE}}$  measured at $M$ angles $\theta_1, \cdots, \theta_M$: 
$
 \mathbf{Y}_k=\left[ \Re(\mathbf{c_{TM}}) \quad \Im(\mathbf{c_{TE}}) \quad \Re(\mathbf{c_{TM}}) \quad \Im(\mathbf{c_{TE}}) \right]^T
$.
Considering all frequencies, vectors $\mathbf{X}=(\mathbf{X}_1,\ldots,\mathbf{X}_K) $ and $\mathbf{Y}=(\mathbf{Y}_1,\ldots,\mathbf{Y}_K) $ respectively define complete system state and observation.

\subsection{Observation model}
The 2D-axisymetric Maxwell solver software ($\mathcal{F}_{2D}$) can predict the observation from the system state. Assuming a multidimensional Gaussian measurement uncertainty model, it leads to the following likelihood model $\mathbf{Y}_k \mid \mathbf{X}_k \sim \mathcal{N}(\mathcal{F}_{2D}(\mathbf{X}_k),\mathbf{R}_k)$ (at a given frequency $f_k$), where $\mathbf{R}_k$ is the covariance matrix (assumed known). To avoid numerous and heavy $\mathcal{F}_{2D}$ computations, we have developed the following global approach.  First, the high dimension state space and the associated system response are explored randomly around expected properties (prior knowledge), computations being massively distributed on HPC machines. 

Let $\mathcal{B}_A={\{(\mathbf{X}_k^{(1)},\mathbf{Y}_k^{(1)}) ,\cdots, (\mathbf{X}_k^{(N_E)},\mathbf{Y}_k^{(N_E)})\}}$  the training data composed of $N_E$ couples. It is then processed by N-D statistical techniques; sensibility analysis and model reduction techniques can possibly reduce the state space dimension. Applying multidimensional regression, it turns out that the model is approximatively linear. According to the studied cases, the  linearity errors, evaluated by residue analysis and bootstrap techniques, are significant but much lesser than the RCS measurement uncertainties. Finally, the following linear Gaussian model can be considered (at a given frequency $f_k$):   
\begin{align}
& \mathbf{Y}_k \mid \mathbf{X}_k \sim \mathcal{N}(\mathbf{A}_k\cdot\mathbf{X}_k+\mathbf{Y}^0_k,\mathbf{R}_k)  \label{mod-obs}
\end{align}
where the deterministic part of the linear model is given by the learned matrix $4M\times4N$ $\mathbf{A}_k$ and the vector $\mathbf{Y}^0_k$.

\subsection{Prior model at a fixed frequency}

Let us fix a frequency $f_k$. We model our a priori knowledge on $\mathbf{X}_k$ with a Gaussian distribution $\mathcal{N} \left( \mathbf{m}_k , \mathbf{P}_k \right)$. The object is divided in $N_b$ blocks of areas, each of them composed of a rather homogeneous material. The location of these blocks is known exactly. Prior mean value $\mathbf{m}_k$ is defined with reference values of $\epsilon^{\prime}$, $\epsilon^{\prime \prime}$, $\mu^{\prime}$ and $\mu^{\prime \prime}$ for each of theses blocks. Then, for any component $\mathbf{X}_k^i$, we define a variance ${\sigma_{S,k}^i}^2$ as a mix between absolute and relative uncertainty. To take into account the spatial local homogeneity, covariance matrix $\mathbf{P}_k$ is defined block by block independently (and separately for each of the $\epsilon^{\prime}$, $\epsilon^{\prime \prime}$, $\mu^{\prime}$, $\mu^{\prime \prime}$) by correlation relations between $\mathbf{X}_k$'s block-sharing components:
$$
Cov(\mathbf{X}_k^i,\mathbf{X}_k^j) = {\sigma_{k,S}^i}^2 \cdot \rho_S^{|i-j|}
$$

where $\rho_S \in [0,1]$ is a spatial correlation parameter (typically $\rho_S = 0.95$ in our applications). It means that the correlation between $2$ areas dicreases geometrically at speed $\rho_S$ with the distance between them.

\subsection{Inter-frequential prior model}

Radioelectric properties $\epsilon^{\prime}$, $\epsilon^{\prime \prime}$, $\mu^{\prime}$, $\mu^{\prime \prime}$ are known to vary in function of the wave frequency \cite{Knott};  their  non-stationary dynamic can be quite different from frequency $f_1$ to $f_K$. However, in order to take account of expected frequency profiles regularity (for each material and $\epsilon^{\prime}$, $\epsilon^{\prime \prime}$, $\mu^{\prime}$, $\mu^{\prime \prime}$), we model sequence $(\mathbf{X}_k, k \in \{1, \ldots , K \}$ with a linear Gaussian correlation structure, given by the generalised autoregressive (AR) random process \cite{Pandit}:

\begin{align}
 & \mathbf{X}_1 \sim \ \mathcal{N} \left( \mathbf{m}_1, \mathbf{P}_1\right)  \nonumber \\
 & \left( \mathbf{X}_{k+1} - \mathbf{m}_{k+1} \right) = \mathbf{D}_{\rho} \cdot \mathbf{H}_{k+1} \cdot \mathbf{H}_k^{-1} \cdot \left( \mathbf{X}_k - \mathbf{m}_k \right) + \sqrt{\mathbf{I_d}-\mathbf{D}_{\rho}^2} \cdot \mathbf{H}_{k+1} \cdot \mathbf{V}_k  \label{mod-AP}
\end{align}

where
\begin{itemize}
 \item for each $k$, $\mathbf{H}_k$ is the ``square root'' matrix of covariance matrix $\mathbf{P}_k$, defined as being the unique symetric definite positive matrix satisfying $\mathbf{H}_k \cdot \mathbf{H}_k^T = \mathbf{P}_k$;
 \item $(\mathbf{V}_k,k \in \{1, \ldots , K \})$ are i.i.d. $\mathcal{N} (0,\mathbf{I_d})$;
 \item $\mathbf{D}_{\rho}$ is a positive diagonal matrix commuting with the $\mathbf{H}_k$, further described, and depending on a parameter $\rho$.
\end{itemize}
One notices that so defined sequence $\mathbf{X}_k$ still admits $\mathcal{N} \left( \mathbf{m}_k , \mathbf{P}_k \right)$ as marginal distributions. More generally, it can be shown that the distribution of concatenated vector $\mathbf{X}=(\mathbf{X}_1,\ldots,\mathbf{X}_K)$ is a Gaussian distribution with mean $\mathbf{m}=(\mathbf{m}_1,\ldots,\mathbf{m}_K) $ and covariance matrix:
$$ 
\mathbf{P}  =  \mathbf{\mathcal{H}} \cdot
\left( \begin{array}{ccccc}
\mathbf{I_d} & \mathbf{D}_{\rho} & \mathbf{D}_{\rho}^2 & \cdots & \mathbf{D}_{\rho}^{K-1}  \\
\mathbf{D}_{\rho} & \mathbf{I_d}  & \mathbf{D}_{\rho} &  &  \vdots \\
\mathbf{D}_{\rho}^2 & \mathbf{D}_{\rho} & \ddots & \ddots & \vdots \\
\vdots &  &  \ddots & \ddots &  \mathbf{D}_{\rho}  \\
\mathbf{D}_{\rho}^{K-1} & \cdots & \cdots & \mathbf{D}_{\rho} & \mathbf{I_d}
                        \end{array} \right)
\cdot \mathbf{\mathcal{H}}^T
$$

where $\mathbf{\mathcal{H}}$ is the diagonal by block matrix $\mathbf{\mathcal{H}}=\textmd{diag} (\mathbf{H}_1, \ldots, \mathbf{H}_K) $, which explicits every joint distribution $(\mathbf{X}_i , \mathbf{X}_j)$.\\
Finally, let us clarify parameter $\rho$ (and matrix $\mathbf{D}_{\rho}$). $\mathbf{D}_{\rho}$'s role in these equations makes clear that it's a frequential correlation parameter. Its dimension is to be chosen among $3$ possibilities according to our assumptions.
\begin{itemize}
 \item \underline{$1^{\textmd{st}}$ case}: $\rho$ is assumed independant from material and $\epsilon^{\prime}$, $\epsilon^{\prime \prime}$, $\mu^{\prime}$, $\mu^{\prime \prime}$. It is one-dimensional, $\rho \in [0,1]$, and $\mathbf{D}_{\rho} = \rho.\mathbf{I_d}$.
 \item \underline{$2^{\textmd{nd}}$ case}: $\rho$ depends on the material (block). It is $N_b$-dimensional, $\rho \in [0,1]^{N_b}$, and $\mathbf{D}_{\rho}$  giving to each line (component of $\mathbf{X}_k$) its associated component of $\rho$.
 \item \underline{$3^{\textmd{rd}}$ case}: $\rho$ depends on material and $\epsilon^{\prime}$, $\epsilon^{\prime \prime}$, $\mu^{\prime}$, $\mu^{\prime \prime}$. It is $4.N_b$-dimensional, and $\mathbf{D}_{\rho}$ is the diagonal matrix composed with $4 .N_b$ $\rho_i . \mathbf{I_d}$-type blocks.
\end{itemize}

\subsection{Global model}

Conditionnaly to frequential correlation parameter $\rho$, the problem of the determination of  $\mathbf{X} = (\mathbf{X}_1,\cdots,\mathbf{X}_K)$ given the measurements $\mathbf{Y} = (\mathbf{Y}_1,\cdots,\mathbf{Y}_K)$ can be expressed as a classic linear Gaussian hidden dynamic Markov process observed at "times" $f_k$ ($k=1,\cdots,K$):
\begin{equation*} 
\mathbf{X}_{k+1}=\mathbf{M}_{k}^{\rho}\cdot\mathbf{X}_{k}+\mathbf{w}^{(1)}_k \quad \mbox{and}  \quad \mathbf{Y}_{k}=\mathbf{A}_{k}\cdot\mathbf{X}_{k}+\mathbf{w}^{(2)}_k
\end{equation*}
where $\mathbf{w}^{(1)}_k$ and $\mathbf{w}^{(2)}_k$ are independant Gaussian noises with known parameters, and $\mathbf{M}_{k}^{\rho}$ a known matrix depending on $\rho$ (see (\ref{mod-obs}) and (\ref{mod-AP})). Parameter $\rho$, intuitively representing the inter-frequency regularity, is unknown and to be estimated. In respect to the probabilistic point of vue, it is probabilized, and given a prior distribution $p(\rho)$.


\section{Sequential Monte Carlo approach for global inversion}

\subsection{Rao-Blackwellised SMC algorithm}

The posterior distribution $p(\mathbf{X},\rho| \mathbf{Y})$ can be decomposed as: $ p(\mathbf{X},\rho| \mathbf{Y}) =  p(\mathbf{X}| \rho,\mathbf{Y}) \cdot p(\rho|\mathbf{Y})$. Conditionally to $\rho$, the system is linear Gaussian: the conditional distributions $p(\mathbf{X}| \rho,\mathbf{Y})$ can be straightforwardly computed by Kalman filtering, including in this off-line context backward Kalman smoothing. On the other hand, the term $p(\rho|\mathbf{Y}) \propto p(\mathbf{Y}| \rho) \cdot p(\rho)$ can be evaluated (up to a normalising constant) for a given $\rho$ using the likelihood term provided by the Kalman filter and prior distribution $p(\rho)$. Consequently, in order to exploit this conditional structure of the system, Kalman smoothers are applied and integrated in an interacting particle approach. This idea of mixing analytic integration (here Kalman evaluation of $p(\mathbf{X}|\rho, \mathbf{Y})$) with stochastic sampling is a variance reduction approach, known as Rao-Blackwellisation \cite{Liu}. 

Similarly to \cite{Liu}, we choose to implement an efficient interacting particle  approach, in order to estimate the marginal distribution $ \eta(\rho):= p(\rho | \mathbf{Y}) \propto p(\mathbf{Y}| \rho) \cdot p(\rho)$. Sequential Monte Carlo (SMC) is a stochastic algorithm to sample from complex high-dimensional probability distributions. The principle (e.g., \cite{DMDJ}) is to approximate a sequence of target probability distributions $(\eta_n)$ by a large cloud of random samples termed particles $(\zeta_{n}^k)_{1 \leq k \leq N_p} \in E^{N_p}$, $E$ being called the state space. Between ``times'' $n-1$ and $n$, the particles evolve in state space $E$ according to $2$ steps:
\begin{itemize}
 \item a \textbf{selection} step: every particle $\zeta_{n-1}^{i}$ is given a weight $\omega_i$ defined by a selection function $g_n$ ($\omega_i = g_n(\zeta_{n-1}^{i})$). By resampling (stochastic or deterministic), low-weighted particles vanish and are replaced by replicas of high-weighted ones.
 \item a \textbf{mutation} step: each selected particle $\widehat{\zeta}_{n-1}^{i}$ move, independently from the others, according to a Markov kernel $M_n$.
\end{itemize}

\[
\left.
\begin{array}
[c]{cl}%
\zeta_{n-1}^{1} & \\
\vdots & \\
\zeta_{n-1}^{i} & \\
\vdots & \\
\zeta_{n-1}^{N_p} &
\end{array}
\right] \underbrace{ \overset{g_n} %
{-\!\!\!\!-\!\!\!\!-\!\!\!\!-\!\!\!\!-\!\!\!\!-\!\!\!\!-\!\!\!\!-\!\!\!\!-\!\!\!\!\longrightarrow
} }_{\textmd{selection}}\left[
\begin{array}
[c]{cl}%
\widehat{\zeta}_{n-1}^{1} & \overset{M_{n}}%
{-\!\!\!\!-\!\!\!\!-\!\!\!\!-\!\!\!\!-\!\!\!\!-\!\!\!\!-\!\!\!\!-\!\!\!\!-\!\!\!\!\longrightarrow
}\\
\vdots & \\
\widehat{\zeta}_{n-1}^{i} &
-\!\!\!\!-\!\!\!\!-\!\!\!\!-\!\!\!\!-\!\!\!\!-\!\!\!\!-\!\!\!\!-\!\!\!\!-\!\!\!\!\longrightarrow
\\
\vdots & \\
\widehat{\zeta}_{n-1}^{N_p} &
\underbrace{-\!\!\!\!-\!\!\!\!-\!\!\!\!-\!\!\!\!-\!\!\!\!-\!\!\!\!-\!\!\!\!-\!\!\!\!-\!\!\!\!\longrightarrow}_{\textmd{mutation}}
\end{array}
\right.  \left.
\begin{array}
[c]{cl}%
\zeta_{n}^{1} & \\
\vdots & \\
\zeta_{n}^{i} & \\
\vdots & \\
\zeta_{n}^{N_p} &
\end{array}
\right]
\]

Evolving this way, the cloud of particles, and more precisely the occupation distribution $\eta_n^{N_p} = \frac{1}{N_p} \sum_{k=1}^{N_p} \delta_{\zeta_k^n}$ (sum of Dirac distributions), approximates for each $n$ the theoretical distribution $\eta_n$ defined recursively by the Feynman-Kac formulae, associated with the potentials $g_n$ and kernels $M_n$ \cite{DM}.\\
Back to our objective of sampling from $\eta$, we then define the sequence of distributions $\eta_n(\rho)$:

$$ \eta_n(\rho) \propto p(\mathbf{Y}| \rho)^{\alpha_n} \cdot p(\rho) $$

where  $(\alpha_n)_{0 \leq n \leq n_f}$ is a sequence of number increasing from $0$ to $1$, so that: $\eta_0$ is prior distribution $p(\rho)$, easy to sample, $\eta_{n_f}$ is target distribution $\eta$ and  sequence $(\eta_n)$ admits a Feynman-Kac type structure with calculable selection functions $g_n$ and Markov kernels chosen so that $\eta_n.M_n = M_n$ (Metropolis-Hastings for example). The distribution $\eta$ is then interpreted as being the last distribution of a Feynman-Kac sequence, on which SMC can be performed, the estimator of $\eta$ being the occupation distribution $\eta^{N_p}$ extracted from the last cloud of particles.

\subsection{Results}

\begin{figure}[h]
\centering{\includegraphics[width=1\columnwidth]{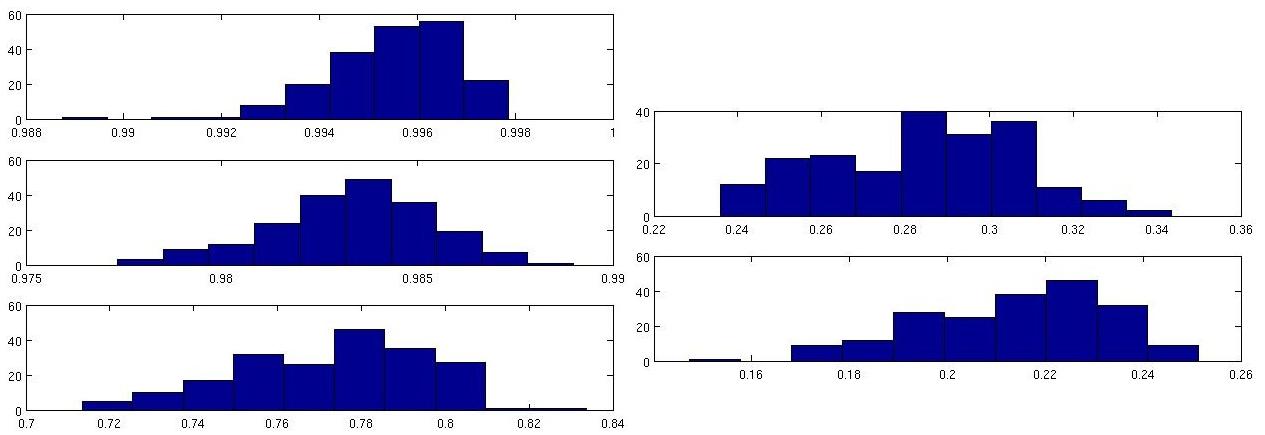}
\caption{\label{hist-rau}Estimation of a $5$-dimensional $\rho$'s distribution with $N_p=100$ particles.}}
\end{figure}

The occupation distribution $\eta^{N_p}$ which approximates $\eta = p(\rho | \mathbf{Y}) $ can be represented dimension by dimension via histograms (see figure \ref{hist-rau}).

\begin{figure}[h]
\centering{\includegraphics[width=0.8\columnwidth]{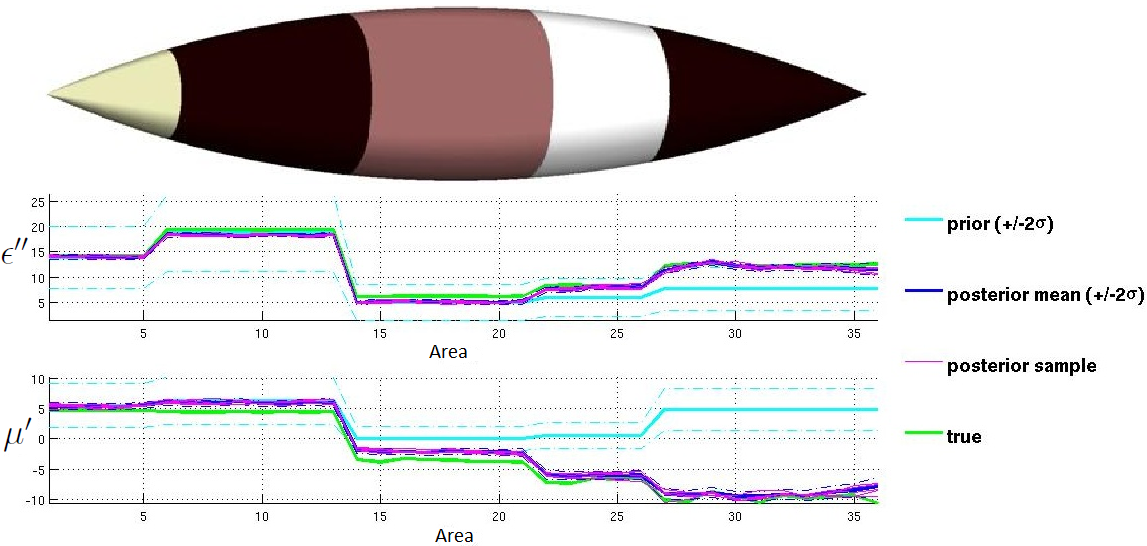}
\caption{\label{freq-fix}Estimated radioelectric properties of a $N_b=5$-block object ($N=19$ areas), at fixed frequency $f_k$.}}
\end{figure}

For each frequency $f_k$, this approximation $\eta^{N_p} \simeq \eta$, associated with the theoretical conditioning relations
\begin{align*}
 &  E(\mathbf{X}_k | \mathbf{Y})    =     E\left[ E(\mathbf{X}_k | \rho,\mathbf{Y}) | \mathbf{Y} \right]  \\
 & Cov(\mathbf{X}_k | \mathbf{Y}) =  E\left[ Cov(\mathbf{X}_k | \rho,\mathbf{Y})  | \mathbf{Y} \right]
+ Cov\left[ E(\mathbf{X}_k | \rho,\mathbf{Y})  | \mathbf{Y} \right]
\end{align*}

can deliver estimators of respectively the mean and the covariance matrix of $\mathbf{X}_k$ .Roughly speaking, the posterior estimation is  performed  by randomly picking a $\rho_i$  from the final cloud of particles and computing associated samples of $\mathbf{X}_k$  by a Kalman smoother conditionally to $\rho_i$. It is illustrated in figure \ref{freq-fix}, with a good agreement between the true state and estimated state.

\begin{figure}[h]
\centering{\includegraphics[width=0.8\columnwidth]{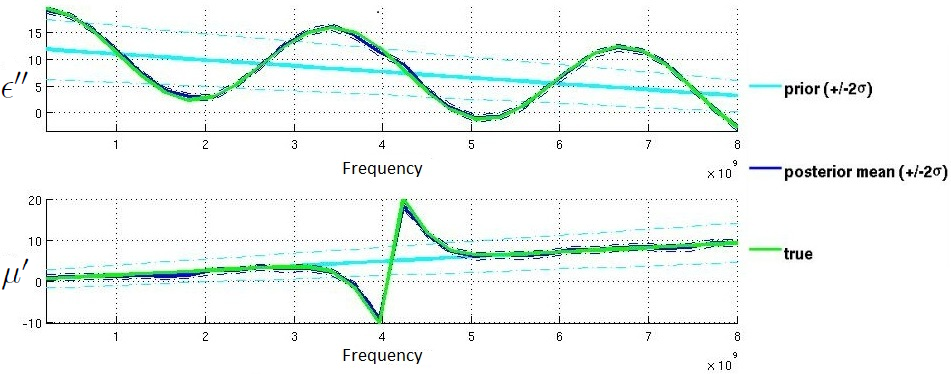}
\caption{\label{ann-fix}Estimated radioelectric properties ($K=30$ frequencies) of a fixed area.}}
\end{figure}

Moreover, for any fixed area, the method provides estimators of the mean and marginal variance for every frequency, so that the results can be presented as frequential profiles, with marginal uncertainties (see figure \ref{ann-fix}). Even when the true (simulated) EM property profiles are chosen markedly divergent from the prior AR-type model, the method turns out to be robust.  It results from the adaptive estimation of $\rho$ which provides small values of $\rho$  (i.e. weak correlation of EM properties for close frequencies) in the case of highly irregular true profiles.

\section{Conclusion}
An efficient statistical inference approach has been applied to estimate local material radioelectric properties from  global EM scattering measurements. It combines intensive computations, meta-modeling and advanced sequential Monte Carlo techniques dedicated to frequency dynamic estimation.

\end{document}